\documentclass[11pt]{article}
\usepackage[dvips]{graphicx}

\begin{document}

\date{\empty}

\title{\textbf{Origin of cosmic magnetic fields: Superadiabatically amplified modes in open Friedmann universes}}

\author{J.D. Barrow${}^1$, C.G. Tsagas${}^{2,3}$ and K. Yamamoto${}^1$\\ {\small ${}^1$DAMTP, Centre for Mathematical Sciences, University of Cambridge}\\ {\small Wilberforce Road, Cambridge CB3 0WA, UK}\\ {\small ${}^2$Section of Astrophysics, Astronomy and Mechanics, Department of Physics}\\ {\small Aristotle University of Thessaloniki, Thessaloniki 54124, Greece}\\ {\small ${}^3$NORDITA, AlbaNova University Center, Roslagstullsbacken 23, SE-10691 Stockholm, Sweden}}

\maketitle

\begin{abstract}
Cosmological magnetic fields in open Friedmann universes can experience superadiabatic amplification within the realm of conventional electromagnetism. This is possible mathematically, despite the conformal invariance of Maxwell's equations, because Friedmann spacetimes with non-Euclidean spatial geometry are not globally conformal to Minkowski space. Physically, this means that even universes that are marginally open today can sustain large-scale magnetic fields that are substantially stronger than previously anticipated. In the present article, we investigate this purely geometric amplification mechanism in greater detail, focusing on the early evolution of the electromagnetic modes in inflationary Friedmann models with hyperbolic spatial geometry. This also allows us to refine the earlier numerical estimates and provide the current spectrum of the residual, superadiabatically amplified magnetic field.
\end{abstract}

\section{Introduction}
Large-scale magnetic ($B$) fields have been observed throughout the universe: from galaxies and galaxy clusters to remote protogalactic structures~\cite{K}. Nevertheless, the origin of these fields is still an open question. The galactic dynamo can explain the galactic and (possibly) the cluster $B$-fields, but the mechanism faces difficulties with those seen in high-redshift protogalaxies~\cite{KA}. The problem deepens when one takes into account recent claims for the detection of magnetic fields in empty intergalactic space~\cite{TGFBGC}, where the dynamo amplification presumably cannot operate. Cosmology could provide the answer, but generating sustainable $B$-fields in the early universe has been proved very challenging~\cite{GR}. The main obstacle is that conventional magnetic fields in spatially flat Friedmann-Robertson-Walker (FRW) universes, which provide the standard cosmological models, are drastically diluted by the universal expansion, especially during inflation. This leads to residual $B$-fields that are far too weak to seed and sustain the galactic dynamo. For decades, the solution has been usually sought outside Maxwellian electromagnetism, standard cosmology, or beyond the linear regime. We can do this in many ways, which explains the number and the variety of the proposed scenarios (see~\cite{KCMW} for relatively recent work and~\cite{GR} for more references). Nevertheless, one could still produce appreciable magnetic seeds of cosmological origin at the linear level and within conventional physics, by appealing to a purely general relativistic (geometrical) amplification mechanism that operates when the FRW background universe has negatively curved spatial sections~\cite{TK}.

Electromagnetic fields are, so far, the only known energy sources of vector nature. This guarantees a purely geometrical coupling between the Maxwell field and the geometry of the host spacetime, ensuring a special status for electromagnetism in Einstein's theory. The gravito-electromagnetic interaction is monitored by the Ricci identities and adds to the standard interplay between matter and geometry of the field equations. Technically, it ensures the presence of curvature-related terms in the electromagnetic wave equations~\cite{E}. Physically, this implies a very different evolution for the Maxwell field in curved spacetimes. Thus, electromagnetic fields in Friedmann models with nonzero spatial curvature evolve differently than in their spatially flat counterpart. More specifically, in FRW universes with negative 3-curvature the adiabatic magnetic decay slows down and the field is superadiabatically amplified~\cite{TK}. This is possible, despite the conformal invariance of standard electromagnetism and the conformal flatness of the FRW spacetimes, because Friedmann models with non-Euclidean spatial geometry are only\textit{locally} conformal to Minkowski space (e.g.~see~\cite{S}). Thus, the aforementioned amplification occurs on relatively large scales, where the curvature effects are prominent and the conformal mapping between the Friedmann and the Minkowski spacetimes breaks down. The mechanism seems to work in all negatively curved FRW models, irrespective of their matter content and, in principle, can lead to residual magnetic fields strong enough to seed the galactic dynamo. As yet, this is the only scenario where superadiabatic magnetic amplification is achieved at the classical level and without introducing any new physics.

The purely geometrical mechanism of magnetic amplification outlined above, has been discussed and developed within the framework of the Friedmann models in~\cite{TK}. It applies to more general spacetimes, however, since analogous exact results have also been obtained in Bianchi class B models that contain electromagnetic fields~\cite{Y}. Overall, the amplification effect appears to be a generic feature of cosmological models with negative spatial curvature. As a result, the latter can sustain large-scale magnetic fields much stronger than generally expected. More specifically, the final strength of the amplified $B$-field can reach magnitudes close to $10^{-15}$~G today, depending on the precise model of inflation~\cite{TK}. These numerical estimates correspond to magnetic modes with sizes close to the curvature scale of an open FRW universe. In particular the aforementioned studies followed the evolution of the mode from the moment it crossed outside the Hubble radius, during a period of slow-roll inflation, throughout the subsequent radiation and dust epochs, all the way to the present. Here, we will turn our attention to the early stages of the magnetic evolution prior and during the slow-rolling phase. Our aim will be to examine whether the superadiabatically amplified magnetic modes can be causally connected at the onset of inflation and the time they crossed the Hubble horizon during the inflationary expansion. Causality will ensure the physical existence of these modes, at least in classical (purely general relativistic) terms. The time of first horizon crossing, on the other hand, will essentially determine the modes' subsequent amplification and eventually their residual strength today. We then provide the spectrum of the present amplitude of the superadiabatically amplified magnetic modes, both analytically and numerically. According to our results, for reasonable values of the cosmological parameters, the final $B$-field lies between $10^{-20}$ and $10^{-12}$ Gauss, which means that it falls fairly comfortably within the typical galactic-dynamo requirements.

\section{Magnetic fields on open FRW backgrounds}
Cosmological magnetic fields are thought to decay adiabatically in all FRW backgrounds at all times. The reason is believed to be the conformal invariance of classical electromagnetism, combined with the conformal flatness of the Friedmann models. Nevertheless, this is not the case when the FRW host has hyperbolic spatial geometry.

\subsection{The gravito-electromagnetic interaction}
Electromagnetic fields are, as yet, the only known energy source of vector nature. This means that the Maxwell and the gravitational fields have a twofold interaction. The first is the familiar interplay between matter and geometry introduced by Einstein's equations. The second is a purely geometrical coupling, which holds even outside the realm of general relativity and is monitored by the Ricci identities. When applied to the magnetic vector ($B_{a}$), the latter read
\begin{equation}
2\nabla_{\lbrack a}\nabla_{b]}B_{c}= R_{abcd}B^{d} \hspace{10mm} \mathrm{and} \hspace{10mm} 2\mathrm{D}_{[a}\mathrm{D}_{b]}B_{c}= \mathcal{R}_{abcd}B^{d}\,,  \label{Ricci}
\end{equation}
where the former applies to the whole spacetime and the latter holds on the observer's 3-dimensional (irrotational) rest-space. Note that $\nabla_{a}$ and $\mathrm{D}_{a}$ are the 4-D and 3-D covariant derivative operators respectively; $R_{abcd}$ and $\mathcal{R}_{abcd}$ are the corresponding Riemann tensors.

This interaction between gravity and electromagnetism implies that the evolution of large-scale cosmological magnetic fields is affected by the geometry of their host spacetime. More specifically, the use of the Ricci identities adds curvature-related terms to the wave equation of the electromagnetic field~\cite{E}. In particular, on an FRW background, the magnetic component of a source-free electromagnetic field is monitored by
\begin{equation}
\ddot{B}_{a}- \mathrm{D}^{2}B_{a}= -5H\dot{B}_{a}- 2\left(\dot{H}+3H^{2}+a^{-2}K\right)B_{a}\,,  \label{ddotBa} \end{equation}
to linear order. Here, $H=\dot{a}/a$ is the background Hubble parameter ($a=a(t)$ is the associated metric expansion scale factor), $K=0,\pm1$ is the 3-curvature index of the underlying FRW model, $\mathrm{D}^{2}=\mathrm{D}^{a}\mathrm{D}_{a}$ is the covariant Laplacian operator of the spatial hypersurfaces and overdots denote comoving proper-time derivatives. Note the gravito-magnetic term on the right-hand side of Eq.~(\ref{ddotBa}). It results from the magneto-geometrical interactions manifested in the Ricci identities (particularly in (\ref{Ricci}b)) and will play the key role in the analysis that follows.

We can simplify expression (\ref{ddotBa}) by introducing the rescaled magnetic field vector $\mathcal{B}_{a}=a^{2}B_{a}$ and by using conformal, instead of proper, time (i.e.~$\eta $, with $\dot{\eta}=1/a$). Then, Eq.~(\ref{ddotBa}), written for the $n$-th magnetic mode, reduces to
\begin{equation}
\mathcal{B}_{(n)}^{\prime\prime}+ (n^{2}+2K)\mathcal{B}_{(n)}=0\,, \label{ddotcBn}
\end{equation}
where $n$ is the eigenvalue of the Laplacian that represents the comoving wavenumber of the mode and primes denote conformal-time derivatives~\cite{TK}. On FRW backgrounds with Euclidean spatial geometry (i.e.~when $K=0$), one can easily show that the rescaled magnetic field remains constant in terms of conformal time. This ensures an adiabatic decay for all magnetic modes (i.e.~$B_{(n)}\propto a^{-2}$, with $n\geq 0$) at all times. Following (\ref{ddotcBn}), for modes with $n^{2}\gg2$, the adiabatic decay-rate persists in Friedmann models with nonzero 3-curvature as well. This is to be expected, since the aforementioned modes correspond to small scales where the curvature effects are unimportant. On large enough lengths, however, the adiabatic decay law is no longer guaranteed due to the magneto-geometrical term on the right-hand side of (\ref{ddotcBn}).\footnote{The Minkowski-like evolution of the rescaled $\mathcal{B}$-field on flat Friedmann models can also be seen as the direct consequence of the conformal invariance of standard electromagnetism and of the fact that the spatially flat FRW spacetime is globally conformal to the Minkowski space. The latter does not apply to Friedmann models with nonzero 3-curvature, which are only locally conformal to the Minkowski space~\cite{S}. In particular, the conformal mapping between the two spacetimes breaks down on large enough scales where the curvature effects become prominent. It is on these wavelengths that the magneto-geometrical term in Eq.~(\ref{ddotcBn}) takes over and the adiabatic magnetic decay law is no longer guaranteed, despite the conformal invariance of standard electromagnetism.} As mentioned earlier, the latter reflects the purely general relativistic interaction between magnetism and spacetime curvature monitored by the Ricci identities.

\subsection{Superadiabatic magnetic amplification}\label{ssSMA}
During a period of inflationary expansion the universe is believed to behave like a very poor electrical conductor. After inflation and reheating, the conductivity is high and the resulting electric currents will freeze any large-scale magnetic field that may be present into the primordial plasma. In highly conductive environments the $B$-field drops adiabatically at all times (i.e.~$B_{a}\propto a^{-2}$) and irrespective of the background geometry. On scales lying beyond the Hubble radius, however, causality ensures the absence of currents. There, the conductivity remains low and the ideal magnetohydrodynamic (MHD) approximation does not hold. In such an environment, the magnetic field vector is still monitored by Eqs.~(\ref{ddotBa}) and (\ref{ddotcBn}).

According to (\ref{ddotcBn}), the adiabatic magnetic decay persists on closed FRW backgrounds (where $K=+1$). There, the magneto-curvature term simply modifies the frequency of the field's oscillation. On open FRW backgrounds, however, the same term can change the nature of the magnetic wave equation. In particular, setting $K=-1$ in (\ref{ddotcBn}), the latter reads
\begin{equation}
\mathcal{B}_{(n)}^{\prime\prime}+ (n^{2}-2)\mathcal{B}_{(n)}= 0\,,
\label{ddotcBn-1}
\end{equation}
with $n^{2}\geq0$. Therefore, on large enough scales (those with $-2\leq n^{2}-2<0$), the solution of Eq.~(\ref{ddotcBn-1}) no longer leads to conventional wave solutions, but to `exponential waves' of the form
\begin{equation}
\mathcal{B}_{(n)}= \mathcal{C}_{1}\cosh \left(\eta\sqrt{2-n^{2}}\right)+ \mathcal{C}_{2}\sinh \left(\eta\sqrt{2-n^{2}}\right)\,,  \label{cBn}
\end{equation}
where now $0\leq n^{2}<2$. When our open FRW background contains a single perfect fluid of barotropic index $w=p/\rho\neq-1/3$, the associated scale factor evolves as $a^{\beta}\propto \sinh(\beta\eta)$, where $\beta=(1+3w)/2$. Then, expression (\ref{cBn}) recasts into the evolution law
\begin{equation}
B_{(n)}= C_{1}\left({\frac{a}{a_{0}}}\right)^{\sqrt{2-n^{2}}-2}+ C_{2}\left({\frac{a}{a_{0}}}\right)^{-\sqrt{2-n^{2}}-2}\,, \label{Bn}
\end{equation}
for the actual magnetic field. The above implies superadiabatic amplification, namely a decay rate slower than the adiabatic one, for all modes with $0\leq n^{2}<2$. Also, given that $n^{2}=1$ corresponds to the curvature scale of the FRW background (see \S~\ref{ssCHs} below), we deduce that the amplification applies to all supercurvature modes (i.e.~those with $0\leq n^{2}<1$), as well as to the largest subcurvature modes (i.e.~the ones with $1\leq n^{2}<2$). More specifically, on the curvature scale the magnetic field drops as $B\propto a^{-1}$, while at the homogeneous limit (i.e.~as $n^{2}\rightarrow 0$) its decay rate slows down further to $B\propto a^{\sqrt{2}-2}$.

So far, we have established that magnetic fields on spatially open FRW backgrounds can be superadiabatically amplified by curvature effects within the realm of classical electromagnetic theory. We have also identified the scales spanned by the affected magnetic modes and found that the amplification effect is essentially independent of the type of matter that fills the universe. This means that large-scale $B$-fields can be superadiabatically amplified during slow-roll inflation, reheating, and throughout the subsequent radiation and dust eras. On physical grounds, we would also like to know whether the scales of interest can be causally connected at the onset of the inflationary expansion, as well as the time they cross outside the Hubble horizon. To large extent, the latter will decide the residual strength of the associated superadiabatically amplified $B$-fields. It will therefore help to investigate how the hyperbolic spatial geometry of the open FRW models affects causality and modifies the standard slow-roll scenario.

\section{Causality and inflation in open FRW models}
When applied to spatially flat Friedmann universes, typical slow-roll inflation leads immediately to the familiar de Sitter phase of exponential expansion. This standard picture changes, however, in models with non-Euclidean spatial geometry, especially during their early stages. In what follows, we will discuss certain aspects of FRW cosmologies with negative 3-curvature.

\subsection{Causal horizons}\label{ssCHs} 
Consider a Friedmann universe with hyperbolic spatial hypersurfaces, that contains a single barotropic medium of energy density $\rho$ and isotropic pressure $p$. The evolution of this model is determined by the set of equations
\begin{equation}
\dot{\rho}= -3H\rho(1+w)\,, \hspace{10mm} \dot{H}= -H^{2}- {\frac{1}{6}}\,\kappa\rho(1+3w)  \label{FRW1}
\end{equation}
and
\begin{equation}
H^{2}= {\frac{1}{3}}\,\kappa \rho+ {\frac{1}{a^{2}}}\,, \label{FRW2}
\end{equation}
where $w=p/\rho$ is the barotropic index and $\kappa=8\pi G$ is the rescaled gravitational constant (with $c=1$). Recalling that $\lambda_{H}=1/H$ is the Hubble horizon, $\Omega=\kappa\rho/3H^{2}$ is the density parameter and defining $\lambda_{K}=a$ as the curvature scale of the universe, we see that Eq.~(\ref{FRW2}) is recast as
\begin{equation}
\lambda_{K}= {\frac{\lambda_{H}}{\sqrt{1-\Omega}}}\, \label{lambdaK}
\end{equation}
with $0<\Omega<1$ for all open FRW models (e.g.~see~\cite{LS}). The latter means that the curvature radius of an open FRW cosmology lies always outside the Hubble radius, with the two scales getting close in the low density limit (i.e.~$\lambda_{K}\rightarrow \lambda_{H}^{+}$ as $\Omega\rightarrow0^{+}$).

Let us now go back to the magnetic case discussed in \S~\ref{ssSMA} and consider a mode with physical wavelength $\lambda_{n}=a/n$. We saw that large-scale $B$-fields are superadiabatically amplified during inflation, reheating, and throughout the subsequent evolution of the universe. Given the large wavelengths of these modes, it is of interest to establish whether or not they can be causally connected at the onset of inflation. Recalling that $\lambda_{K}=a,$ and using (\ref{lambdaK}), we arrive at
\begin{equation}
\lambda_{n}= {\frac{\lambda_{H}}{n\sqrt{1-\Omega}}}\,, \label{lambdan}
\end{equation}
which relates the scale of the magnetic perturbation to that of the Hubble horizon. The above ensures that the scale of the mode in question will lie inside the Hubble radius, and therefore it will be in causal contact, as long as
\begin{equation}
\Omega< 1-{\frac{1}{n^{2}}}\,.  \label{cOmega}
\end{equation}
This condition is readily satisfied by all subcurvature lengths, namely by those with $n^{2}>1$. In other words, scales smaller than the curvature radius of a open FRW spacetime can reside inside the Hubble horizon, provided the universe is sufficiently open. These include the largest subcurvature magnetic modes (i.e.~those with $2>n^{2}>1$), which are superadiabatically amplified (see~\S~\ref{ssSMA}). For example, subcurvature magnetic modes with $n=1.01$ and $n=1.1$ will be \emph{within the Hubble radius} if $\Omega<0.02$ and $\Omega<0.17$, respectively, at the beginning of the inflationary phase. Supercurvature modes can never satisfy condition (\ref{cOmega}). This reflects the fact that the corresponding lengths are always outside the Hubble radius. Nevertheless, in principle, even supercurvature lengths can be causally connected in open FRW models.

When dealing with spatially flat FRW cosmologies containing conventional matter, namely with $w>-1/3$, the Hubble length essentially coincides with the particle horizon of the comoving observers and therefore it effectively defines their observable universe. In those models, super-Hubble (and consequently super-curvature) scales are always causally disconnected. This is not the case, however, in open Friedmann models. There, the Hubble and the particle horizons do not coincide, but the latter is generally larger than the former. For instance, at any given time, the particle horizons of a radiation and a dust-dominated open FRW universe are
\begin{equation}
\lambda_{P}= {\frac{1}{2H\sqrt{1-\Omega}}} \ln\left({\frac{1+\sqrt{1-\Omega}}{1-\sqrt{1-\Omega}}}\right) \hspace{10mm} \mathrm{and} \hspace{10mm} \lambda_{P}= {\frac{1}{H\sqrt{1-\Omega}}} \ln\left({\frac{1+\sqrt{1-\Omega}}{1-\sqrt{1-\Omega}}}\right)\,,  \label{rdlambdaP}
\end{equation}
respectively (see also~\cite{W}). Then, a simple Taylor expansion shows that both particle horizons diverge as the energy density of the universe becomes progressively lower (i.e.~$\lambda_{P}\rightarrow+\infty$ as $\Omega\rightarrow0^{+}$ in agreement with the Milne limit). Close to the Euclidean threshold, on the other hand, one recovers the familiar $\lambda_{P}=1/H$ and $\lambda_{P}=2/H$ expressions of the radiation and the dust eras respectively. The conclusion is that super-Hubble, as well as super-curvature, scales can be causally connected in FRW universes with sufficiently low density.\footnote{Our approach is purely general relativistic and causality is the criterion that decides whether physical processes can operate, or not, within a given region. It has been argued that, quantum mechanically, the superadiabatically amplified modes may not be normalisable~\cite{Ad-RD}. The argument was based on a study adopting open inflation, instead of the standard slow-roll scenario used here. The authors found that the modes were normalisable on small scales, where the 3-curvature effects are negligible, but encountered the usual infinities when they reached lengths where the curvature is strong. Leaving aside the question of open inflation, the aforementioned normalisability issues are not surprising, given the absence of a theory unifying general relativity and quantum mechanics. After all, these are exactly the problems that any future theory of quantum gravity is expected to solve.} It is therefore of physical interest to explore their phenomenological consequences.

\subsection{The inflationary dynamics}\label{ssIDs} 
Introducing conformal time and assuming a period of de Sitter-type inflation (when $p=-\rho$) the system of (\ref{FRW1}) and (\ref{FRW2}) has the parametric solution \begin{equation}
a= a_{0}\left[{\frac{e^{\eta}(1-e^{2\eta_{0}})} {e^{\eta_{0}}(1-e^{2\eta})}}\right]  \label{a-eta}
\end{equation}
and
\begin{equation}
t= t_{0}+ {\frac{a_{0}(1-e^{2\eta _{0}})}{2e^{\eta _{0}}}}\, \ln\left[{\frac{(1+e^{\eta})(1-e^{\eta_{0}})} {(1-e^{\eta})(1+e^{\eta_{0}})}}\right] \,,  \label{t-eta}
\end{equation}
where $\eta<0$ and the zero suffix denotes a given time during inflation. Thus, at the beginning of the inflationary phase, when $\eta\ll0$ and $e^{\eta}\ll1$, the scale factor evolves as $a\propto t$. Towards the end of inflation, on the other hand, we have $\eta\rightarrow0^{-}$ and $e^{\eta}\rightarrow1^{-}$, which ensure that $a\propto e^{t}$. Put another way, the slow-rolling regime in an open FRW model starts with coasting expansion and the exponential phase only occurs at the end, when the effects of curvature have faded away. If we are interested in modes close to the curvature scale, it is important to quantify the effect of the coasting phase.

Further insight can be obtained by evaluating the principal kinematic and dynamic variables. In particular, using primes to denote conformal time derivatives, the Hubble and the deceleration parameters are $H={a^{\prime}/a^{2}}$ and $q=-(1+{H^{\prime}/aH^{2}})$ respectively. Then, on using expressions (\ref{a-eta}) and (\ref{t-eta}), we obtain \begin{equation}
H= {\frac{e^{\eta_{0}}(1+e^{2\eta})} {a_{0}e^{\eta}(1-e^{2\eta_{0}})}}\,, \hspace{10mm} \mathrm{and} \hspace{10mm} q= -{\frac{4e^{2\eta}}{(1+e^{2\eta})^{2}}}\,. \label{Hq2}
\end{equation}
The former combines with Eqs.~(\ref{FRW2}) and (\ref{a-eta}) to give
\begin{equation}
{\frac{1}{3}}\,\kappa\rho= {\frac{4e^{2\eta_{0}}} {a_{0}^{2}(1-e^{2\eta_{0}})^{2}}}\,,  \label{rho}
\end{equation}
which guarantees that $\rho=\rho_{0}=$~constant, as expected. Finally, substituting this result in the right-hand side of (\ref{Hq2}a) and keeping in mind that $\Omega=\kappa\rho/3H^{2}$, we arrive at
\begin{equation}
H= {\frac{1}{2}}\,\sqrt{\frac{\kappa\rho_{0}}{3}}\, {\frac{(1+e^{2\eta })}{e^{\eta}}} \hspace{10mm} \mathrm{and} \hspace{10mm} q= -\Omega\,.  \label{s-rHq2}
\end{equation}
Accordingly, at the onset of inflation (i.e.~when $\eta\rightarrow-\infty$) we have $H\rightarrow+\infty$, $q\rightarrow0^{-}$ and $\Omega\rightarrow0^{+}$. On the other hand, towards the end of the inflationary expansion (i.e.~as $\eta\rightarrow0^{-}$), we find that $H\rightarrow\sqrt{\kappa\rho_{0}/3}$, $\Omega\rightarrow1^{-}$ and $q\rightarrow-1^{+}$. As before, we see that the model enters its accelerating, de Sitter-type, phase only asymptotically at the end of the inflationary regime.

\subsection{Hubble horizon crossing}\label{ssHHC} 
During the coasting phase of the expansion, we have $H\propto t^{-1}$, which means that the Hubble horizon scales like $\lambda_{H}\propto a$. Given that all wavelengths also scale in the same way, modes that were originally inside the Hubble radius will remain so and cross outside only towards the end of inflation, when the curvature effects have faded away and the expansion starts to accelerate. As will see next, the time of horizon crossing is decided solely by the comoving wavelength of the mode in question.

Let us consider a mode with physical scale $\lambda_{n}=a/n$, where $n$ is the associated comoving eigenvalue. On using expressions (\ref{a-eta}) and (\ref{Hq2}a), we obtain
\begin{equation}
{\frac{\lambda_{n}}{\lambda_{H}}}= {\frac{aH}{n}}=  {\frac{1+e^{2\eta}}{n(1-e^{2\eta})}}\,,  \label{lambdas1}
\end{equation}
with $\eta<0$. Thus, a mode will lie inside the Hubble radius (i.e.~$\lambda_{n}/\lambda_{H}<1$) as long as
\begin{equation}
n> {\frac{1+e^{2\eta }}{1-e^{2\eta }}}\,.  \label{ncon1}
\end{equation}
Given that $\eta<0$, the latter guarantees that $n>1$ always. In particular, the right-hand side of the above approaches unity at the beginning of the inflationary phase (when $\eta\rightarrow-\infty$) and starts to diverge as we approach the end of inflation (i.e.~as $\eta\rightarrow0^{-}$). In other words, supercurvature scales (those with $0<n^{2}<1$) are always outside the Hubble horizon.

Taking the conformal time derivative of (\ref{lambdas1}), we find that $(\lambda_{n}/\lambda_{H})^{\prime}= 4e^{2\eta}/n(1-e^{2\eta})^{2}>0$ at all times, which ensures that during the inflationary regime any given wavelength grows faster than the Hubble horizon. This in turn implies that, given enough time, essentially all sub-Hubble scales will eventually cross outside. Following (\ref{ncon1}), at the time of horizon-crossing, namely when $\lambda_{n}/\lambda_{H}=1$ and $\eta=\eta_{HC}$, we find that
\begin{equation}
\eta_{HC}= {\frac{1}{2}}\,\ln\left({\frac{n-1}{n+1}}\right)< 0\,,
\label{etaHC}
\end{equation}
since $n>1$ (see Eq.~(\ref{ncon1}) above). Consequently, all finite wavelengths can exit the Hubble scale before the end of inflation. However, for modes that are initially well inside the Hubble radius (i.e.~with $n\gg1$), horizon crossing occurs only at the end of inflation (i.e.~$\eta_{HC}\rightarrow0^{-}$). Further information can be obtained by recasting the Friedmann equation into the form
\begin{equation}
{\frac{\lambda_{n}}{\lambda _{H}}}= {\frac{1}{n\sqrt{1-\Omega}}}\,, \label{lambdas2}
\end{equation}
with $0<\Omega<1$ in our case. Therefore, at the time when a given wavelength passes through the Hubble threshold, the density parameter of the universe is
\begin{equation}
\Omega_{HC}= 1- {\frac{1}{n^{2}}}\,.  \label{OmegaHC}
\end{equation}
where $n^{2}>1$.\footnote{According to expression (\ref{OmegaHC}), modes with $0\leq n^{2}\leq1$ have $\Omega_{HC}\leq0$. This result simply reflects the fact that the corresponding scales lie outside the Hubble length at all times. Recall that $n^{2}=1$ corresponds to the curvature radius, which in open FRW models is always larger than the Hubble horizon (see \S~\ref{ssCHs} earlier).} Again, we see that most subhorizon modes do not cross the Hubble scale before the value of the density parameter has approached substantially close to unity. Wavelengths spanning scales just inside the curvature radius, with $n=1.01$ and $n=1.1$ for example, cross outside the Hubble horizon at $\Omega_{HC}=0.02$ and $\Omega_{HC}=0.17$ respectively. For smaller lengths, say with $n=2$ and $n=10$, horizon crossing occurs later, when $\Omega_{HC}=0.75$ and $\Omega_{HC}=0.99$ respectively. Finally, it is worth pointing out that a given mode re-enters the Hubble horizon (during the subsequent radiation or matter dominated eras) when $\Omega$ is again equal to $\Omega_{HC}$. For the current Hubble scale, whatever $n$ it possesses, this means that $\Omega_{HC}=\Omega_{0}$, a result that will prove useful later.

\subsection{The number of e-folds}\label{ssNe-fs} 
In a FRW universe, the Hubble scale is related to the density parameter by means of the expression $\lambda_{H}= \sqrt{3\Omega/\kappa\rho}$. Therefore, during inflation, a perturbation that crosses the Hubble horizon has wavelength
\begin{equation}
\lambda_{n}= \lambda_{H}\simeq {\frac{M_{Pl}\Omega_{HC}^{1/2}}{M^{2}}}\,,  \label{lambdaHC}
\end{equation}
with $n^{2}>1$. Note that $\kappa\simeq M_{Pl}^{-2}$ and $\rho\simeq M^{4}=$~constant, where $M_{Pl}$ represents the Planck mass and $M$ the energy scale of inflation in natural units. After horizon crossing, the aforementioned wavelength grows by a factor of $e^{N}$, where $N$ is the number of e-folds between horizon-crossing and the end of the inflationary regime, so then we have
\begin{equation}
(\lambda_{n})_{INF}\simeq e^{N}{\frac{M_{Pl}\Omega_{HC}^{1/2}}{M^{2}}}\,.  \label{lambdaINF}
\end{equation}
Throughout the reheating era, the energy density of the matter decreases as $\rho\propto a^{-3}$. Consequently, by the end of that period, the scale in question has grown further by
\begin{equation}
{\frac{a_{RH}}{a_{INF}}}= \left({\frac{\rho_{INF}} {\rho_{RH}}}\right)^{1/3}\simeq \left({\frac{M}{T_{RH}}}\right)^{4/3}\,,  \label{aRD-aRH}
\end{equation}
since $\rho_{RH}\simeq T_{RH}^{4}$ (in natural units). Putting (\ref{lambdaHC}) and (\ref{lambdaINF}) together, we deduce that at the beginning of the radiation era, the original wavelength is
\begin{equation}
(\lambda_{n})_{RH}\simeq e^{N}{\frac{M_{Pl}\Omega_{HC}^{1/2}} {M^{2/3}T_{RH}^{4/3}}}\,,  \label{lambdaRH}
\end{equation}
with the quantities in the right-hand side given in natural units.

Expression (\ref{lambdaRH}) provides the scale spanned at the end of reheating by a wavelength that crossed the horizon $N$ e-folds before inflation was over. We may also estimate the aforementioned number of e-folds using thermodynamic arguments. More specifically, following~\cite{KT}, we assume that the expansion of the universe proceeds adiabatically (with the exception of the reheating era). Then, the entropy contained within the current Hubble scale, which is estimated close to $10^{88}$, has remained unchanged since reheating. At that time, the entropy inside a region that crossed the horizon $N$ e-folds before inflation ended was
\begin{equation}
S_{RH}\simeq \lambda_{RH}^3T_{RH}^3\simeq e^{3N}{\frac{M_{Pl}^3\Omega_{HC}^{3/2}}{M^2T_{RH}}}\,,  \label{SRH}
\end{equation}
where the last equality derives from Eq.~(\ref{lambdaRH}). Putting $S_{RH}$ equal to $10^{88}$ in the above and recalling that $M_{Pl}\simeq10^{19}$~GeV, we find that a scale of the size of the current observable universe, crossed the Hubble horizon
\begin{equation}
N^{\ast}\simeq 24+ {\frac{2}{3}}\,\ln M+ {\frac{1}{3}}\,\ln T_{RH}- {\frac{1}{2}}\,\ln{\Omega^{\ast}}_{HC}\,,  \label{N*}
\end{equation}
e-folds before the end of inflation, where the $\ast$ denotes quantities associated with the current Hubble scale of the universe.

In general, a wavelength of size $\lambda_{n}$ today crossed the Hubble horizon $N$ e-folds before the end of inflation when $\eta=\eta_{HC}$, with the latter given in Eq.~(\ref{etaHC}). Obviously, wavelengths \emph{smaller} than the present Hubble scale crossed \emph{later} and \emph{larger} scales crossed \emph{earlier}. Then, using (\ref{a-eta}), we have
\begin{equation}
e^{N-N^{\ast}}= \frac{a_{HC}^{\ast}}{a_{HC}}= e^{\eta_{HC}^{\ast}- \eta_{HC}}\frac{1-e^{2\eta _{HC}}}{1-e^{2\eta_{HC}^{\ast}}}\,.  \label{N-N*}
\end{equation}
Finally, substituting (\ref{N*}) into the above, while employing (\ref{etaHC}) and (\ref{OmegaHC}), yields
\begin{equation}
N\simeq 53+ \frac{2}{3}\ln\left(\frac{M}{10^{14}}\right)+ \frac{1}{3}\ln\left(\frac{T_{RH}}{10^{10}}\right)- \frac{1}{2}\ln\left(1-\Omega_{0}\right)- \frac{1}{2}\ln\left(n^{2}-1\right)\,,  \label{N}
\end{equation}
given that $\Omega_{HC}^{\ast}=\Omega_{0}$ (see Eq.~(\ref{OmegaHC}) in \S~\ref{ssHHC}). In the above, which generalises Eq.~(8.45) of~\cite{KT} for FRW models with $\Omega<1$, the quantities $M$ and $T_{RH}$ are measured in GeV. Also, $n$ is related to the physical wavelength at present by $n=(1-\Omega_{0})^{-1/2} (\lambda_{H}/\lambda_{n})_{0}$ (see also (\ref{lambdas2})).\footnote{Substituting the relation $n=(1-\Omega_{0})^{-1/2}(\lambda_{H}/\lambda_{n})_{0}$ to the right-hand side of Eq.~(\ref{N}) and then setting $\Omega_{0}=1$, one can easily recover expression (8.45) in~\cite{KT}.}

Looking at expression (\ref{N}), one can see that the number of e-folds does not scale as $N\propto\ln\lambda_n$, in contrast to the flat-FRW case (compare to Eq.~(8.45) in~\cite{KT}). The difference reflects the fact that, in a open FRW models, the Hubble radius is not constant during inflation (see \S~\ref{ssIDs} earlier). Also, following (\ref{N}), the e-folding number diverges as the wavelength approaches the curvature scale (i.e.~for $n^2\rightarrow1$). However, the singularity is apparent and can be explained by recalling that the supercurvature scales, namely those with $0\leq n^2\leq1$, are always outside the Hubble length. Hence, for the corresponding modes, there is no horizon exit and Eq.~(\ref{N}) does not apply. In their case, the key parameter is the total number of e-folds. Estimating the amplification of supercurvature $B$-modes is not essential for our purposes and goes beyond the scope of the present article. Therefore, from now on, we will focus on magnetic fields spanning the largest subcurvature scales (i.e.~modes with $1<n^2<2$).

\section{The residual magnetic field}\label{sRMF} 
The number of e-folds between the time a certain magnetic mode crossed outside the Hubble radius and the end of inflation, is crucial for the current strength of the residual $B$-field. There are additional factors, however, that can also play a key role.

\subsection{The overall amplification}\label{ssOA} 
In line with the Gibbons-Hawking temperature, the energy density ($\rho_{B}$) stored inside a magnetic mode at horizon crossing, is determined by the dimensionless ratio
\begin{equation}
\left({\frac{\rho_{B}}{\rho}}\right)_{HC}\sim \left({\frac{M}{M_{Pl}}}\right)^{4}\,,  \label{GH}
\end{equation}
where $\rho_{B}\sim B^{2}$ is the energy density of the $B$-field. Magnetic fields coherent on the largest subcurvature scales (and beyond) are superadiabatically amplified as $B\propto a^{-\beta(n)}$, with $\beta(n)=2-\sqrt{2-n^{2}}$. Then, given the constant background density throughout inflation, we have
\begin{equation}
\left({\frac{\rho_{B}}{\rho}}\right)_{INF}\sim e^{-2N\beta(n)}\left(\frac{M}{M_{Pl}}\right)^{4}\,,  \label{rhoBINF}
\end{equation}
at the end of the slow-rolling regime.

During reheating, the effective equation of state of the matter is that of pressure-free dust. This means that $\rho_{RH}= \rho_{INF}(a_{INF}/a_{RH})^{3}$ by the end of that period. At the same time, the large scale magnetic fields are still superadiabatically amplified. Therefore, near the curvature scale, where $B\propto a^{-\beta (n)}$, we have $(\rho_{B})_{RH}= (\rho_{B})_{INF}(a_{INF}/a_{RH})^{2\beta(n)}$. As a result, since $\rho_{INF}\simeq M^{4}$ and $\rho_{RH}\simeq T_{RH}^{4}$, the relative strength of the associated magnetic field at the end of reheating is
\begin{equation}
\left({\frac{\rho_{B}}{\rho_{\gamma }}}\right)_{RH}\sim e^{-2N\beta(n)}\left(\frac{M}{T_{RH}}\right)^{[12-8\beta(n)]/3} \left(\frac{M}{M_{Pl}}\right)^{4}\,.  \label{rhoBRH}
\end{equation}
Recall that $\rho_{RH}\simeq\rho_{\gamma}$ at that time, with $\rho_{\gamma}$ representing the energy density of the radiative component. The magnetic amplification continues throughout the radiation epoch and the subsequent dust era. Therefore, for $B$-fields spanning lengths close to the present curvature scale of the universe, we have $(\rho_{B}/\rho_{\gamma})_{0}= (\rho_{B}/\rho_{\gamma})_{RH}(T_{RH}/T_{0})^{2[2-\beta(n)]}$ today. Also, to obtain the last result, we have taken into account that $\rho_{\gamma}\propto a^{-4}$ and $T\propto1/a$. Combining these with (\ref{N}) and setting $T_{0}\sim10^{-13}$~GeV, gives
\begin{equation}
\left(\frac{\rho_{B}}{\rho_{\gamma}}\right)_{0}\sim 10^{-117+102\sqrt{2-n^{2}}} \left(\frac{M}{10^{14}}\right)^{4\sqrt{2-n^{2}}} \left[(1-\Omega_{0})(n^{2}-1)\right]^{2-\sqrt{2-n^{2}}}\,, \label{rhoB0}
\end{equation}
since $\beta(n)=2-\sqrt{2-n^{2}}$ and $\Omega_{HC}^{\ast}= \Omega_{0}$ (see Eq.~(\ref{OmegaHC}) in \S~\ref{ssHHC}). Finally, recalling that $(\rho_{\gamma})_{0}\sim10^{-51}$~GeV${}^{4}$, we arrive at an expression for the present-day magnetic field
\begin{equation}
B_{0}\sim 10^{-65+51\sqrt{2-n^{2}}} \left({\frac{M}{10^{14}}}\right)^{2\sqrt{2-n^{2}}} \left[(1-\Omega_{0})(n^{2}-1)\right]^{(2-\sqrt{2-n^{2}})/2}\,,
\label{B0}
\end{equation}
for the largest subcurvature modes (i.e.~those with $1<n^{2}<2$). This formula provides, in Gauss units, the present magnitude of a superadiabatically magnetic mode that spans scales close to the curvature radius of an open FRW universe.

\subsection{The final magnetic spectrum}\label{ssFMS} 
According to expression (\ref{B0}), the final strength of the superadiabatically magnetic field depends on the energy scale of the adopted inflationary scenario ($M$), on the current density of the universe ($\Omega_{0}$) and on the mode's scale ($n$). The WMAP-normalised data constrain the current density parameter of the universe, so that approximately $1-\Omega_{0}\leq10^{-2}$. Also, typical inflationary models have energy scales between $10^{14}$~GeV and $10^{17}$~GeV. Thus, setting $1-\Omega_{0}\sim 10^{-2}$ and $M\sim10^{14}$~GeV, the current magnitude of a (subcurvature) magnetic mode with $n=1.01$ is
\begin{equation}
B_{0}\sim 10^{-16}\hspace{5mm}\mathrm{Gauss}\,.  \label{B01}
\end{equation}
Stronger $B$-fields can be obtained by increasing the scale of inflation. For instance, keeping the wavelength of the mode and the current density parameter unchanged and assuming that $M\sim10^{16}$~GeV, we arrive at $B_{0}\sim10^{-12}$~G. Magnetic seeds of such strengths are still too weak to affect outcome of primordial nucleosynthesis~\cite{CST}, or leave an observable signature in the CMB~\cite{CMB}. They can readily sustain the galactic dynamo, however, which typically requires an initial $B$-seed with magnitude in the range between $10^{-20}$~G and $10^{-12}$~G~\cite{KA}. In addition, intriguingly, magnetic fields close to $10^{-16}$~G are very close in strength to those recently reported in empty intergalactic space~\cite{TGFBGC}.\footnote{When all the magnetic amplification takes place during inflation and the $B$-field decays adiabatically ever since, strengths around $10^{-16}$~G today can lead to the so called backreaction problems~\cite{DMR}. These occur because the required inflationary amplification is so strong that the magnetic energy density catches up with that of the driving inflaton. In our scenario there is no backreaction issue. The energy density of the $B$-field is much lower than that of the dominant matter component at all times, namely during inflation, reheating and later during the radiation and the dust epochs. Residual strengths close to $10^{-16}$~G are achieved because the superadiabatic amplification of the field persists throughout its evolution and it is not confined to the inflationary era only.}

\begin{figure}[tbp]
\begin{center}
\includegraphics[height=6.8cm, width=15cm]{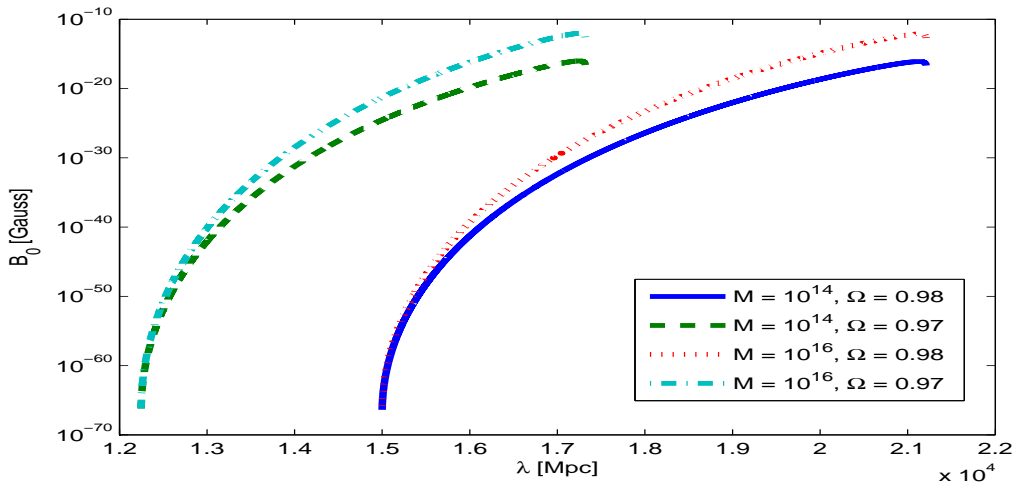}
\end{center}
\caption{The current spectrum of the superadiabatically amplified magnetic field, as a function of the mode wavelength, for different energy scales of inflation, and values of the current density parameter (on a logarithmic scale). Small wavelengths experience weak (or no) amplification, while the maximum enhancement occurs very close to the curvature scale. The cut-off seen at the curvature radius of the universe simply reflects the fact that supercurvature modes are not included in our analysis. Note that, as $\Omega_{0}$ approaches unity, the curvature scale is pushed further away from the Hubble horizon, which is assumed to be at approximately $10^{3}$~Mpc. The dependence of the numerical results on $\Omega_{0}$, however, is considerably weaker than that on $M$.} \label{Fig2}
\end{figure}

The numerical estimates quoted above will drop if the current density parameter gets closer to unity. However, the dependence of the residual magnetic strength on the present density of the universe is relatively weak, and certainly weaker than that on the energy scale of the adopted inflationary model. This means that we can obtain astrophysically relevant $B $-seeds even in very marginally open universes. For example, setting $n=1.01$, $M\sim 10^{14}$~GeV and $1-\Omega_{0}\sim10^{-10}$, we find $B_{0}\sim 10^{-20}$~G. For $M\sim10^{16}$~GeV, the same strength can be obtained even when $1-\Omega_{0}\sim10^{-18}$. These results correspond to $B$-modes coherent on lengths just inside the curvature radius of an open FRW universe, which experience strong superadiabatic amplification. In particular, a mode with $n=1.01$ decays as $B\propto a^{-1.01}$ (see solution (\ref{Bn}) in \S~\ref{ssSMA}). As we move down to smaller scales, however, the magnetic decay-rate increases and at the $n=2$ threshold the adiabatic decay law is re-established (see Fig.~\ref{Fig2}). We also remind the reader that supercurvature lengths, although causally connected, are not included in our analysis (see \S~\ref{ssNe-fs}).

It should be emphasised that the above quoted magnetic strengths do not include the amplification of the field during the nonlinear phase of galaxy formation. In the case of spherically symmetric protogalactic collapse, for example, the magnitude of the magnetic seed typically increases by two or three orders of magnitude. In the more realistic case of anisotropic collapse, shearing effects could add one or two extra orders of magnitude to the $B$-seed~\cite{DBL}. Then, the galactic dynamo is expected to take over~\cite{KA}. Also, our numerical results refer to $B$-fields spanning scales comparable to the curvature radius of the universe. Once galaxy formation begins, however, these fields should start breaking up and reconnecting on scales comparable to that of the collapsing protogalaxy. Note that, on sufficiently small scales, damping effects will process the magnetic spectrum and determine its small-scale structure and polarization~\cite{JKO}.

\section{Discussion}\label{sD}
The problem of cosmic magnetism -- its origin, evolution and implications -- persists. The continual detection of magnetic fields in the universe, makes the search for its solution increasingly pressing.  For cosmologists, the biggest challenge is obtaining sustainable primordial $B$-fields that could successfully seed galactic dynamos. There have been many attempts to solve this problem. With very few exceptions, however, the solution has been sought outside standard electromagnetism, standard cosmology, or general relativity. Nevertheless, the resources of Einstein's theory may not have been exhausted yet and it could still provide the answer. The reason is the geometrical interpretation of general relativity, which gives a special status to the electromagnetic field, due to its vector nature. In practice, this means that, in addition to the Einstein equations, the Maxwell field interacts with the geometry of the host spacetime through the Ricci identities as well. Although this particular coupling between gravity and electromagnetism has been known for some time, it remains under-investigated. Here, we have explored the implications of the aforementioned interaction for the evolution and the survival of large-scale cosmological magnetic fields in greater detail.

We have studied how the generation of seed magnetic fields during standard slow-roll inflation in a marginally open universe can lead to significant superamplification of those seed fields to levels that can provide the initial fields needed for dynamos to produce easily the $B$-fields with magnitudes now observed in galaxies, clusters and intergalactic space. This process of magnetic superamplification relies crucially on the existence of negative spatial curvature so that the primordial magnetic fields do not decay away rapidly at the adiabatic, $B\propto a^{-2}$, rate. Our mechanism appeals to the aforementioned magneto-curvature coupling, which in open universes can significantly slow the decay of the magnetic field, by breaking the global conformal invariance that has often wrongly been assumed to dominate its evolution. The resulting superadiabatic magnetic amplification applies to $B$-fields coherent near and beyond the curvature radius of the universe. Here, we have looked at the largest subcurvature scales, calculated the fields expected for likely parameter choices and displayed the predicted magnetic spectrum (see Fig.~\ref{Fig2}). In all cases, the amplification peaks just inside the curvature scale, which implies a typical correlation length of the order of $10^4$~Mpc for the maximally amplified $B$-field. The latter, however, should break up and reconnect on much smaller (cluster or galactic-size) scales once the nonlinear phase of galaxy formation begins. The residual magnetic strength depends on the energy scale of the adopted inflationary model and on the current curvature radius of the universe. More specifically, setting the energy scale of inflation at $M\sim10^{14}~\mathrm{GeV}^{4}$ and assuming that $1-\Omega_{0}\sim 10^{-2}$, the maximum magnetic strength is approximately $10^{-16}$~G.

These numerical results can be tuned further when the data from the Planck mission narrow down the range of $\Omega_{0}$ and that of the inflation energy scale. Until then, we should underline that in our scenario the earlier inflation starts, the stronger the magnetic amplification. On the other hand, the closer the density parameter gets to unity, the weaker the final $B$-field. Nevertheless, the $\Omega$-dependence is relatively weak, which means that even in (very) marginally open universes the residual magnetic seeds are comfortably within the standard dynamo requirements. Finally, before closing, we should also point out that $B$-fields around $10^{-16}$~G are intriguingly close to those recently reported in empty intergalactic space. Overall, our analysis shows that spacetime curvature can act as an effective dynamo mechanism of purely geometrical nature.\newline

\section*{Acknowledgments}
C.T. would like to thank the Institute of Astrophysics at NORDITA and its members for their support and hospitality during his stay in Stockholm, where a large part of this work took place. He also wishes to acknowledge support from DAMTP and STFC at Cambridge University. K.Y. acknowledges the support from the Aristotle University of Thessaloniki and would like to thank Joanna Hatziantonaki for her help with the computing there. K.Y. is supported by the Cambridge Overseas Trust.

\end{document}